\documentclass[10pt]{iopart}

\usepackage{cite}

\usepackage{graphicx}                                               
\usepackage{epstopdf}                                               
\usepackage{color}                                                     
\usepackage{bm}                                                        
\usepackage{appendix}                                              
\usepackage[utf8]{inputenc}
\usepackage{bbold}
\usepackage{bbm}
\usepackage{ulem}
\usepackage{latexsym}
\usepackage[colorlinks=true,citecolor=blue,linkcolor=magenta]{hyperref}
\usepackage{subfigure}

\def\be{\begin{equation}}
\def\ee{\end{equation}} 
\def\bea{\begin{eqnarray}}
\def\eea{\end{eqnarray}}
\def\ra{\rangle}
\def\la{\langle}
\def\bi{\begin{itemize}}
\def\ei{\end{itemize}}
\def\ben{\begin{enumerate}}
\def\een{\end{enumerate}}

\definecolor{dgreen} {RGB}{78,138,21}

\definecolor{orange} {RGB}{255,120,0}

\begin{document} 

\title{Topological Time Crystals}

\author{Krzysztof Giergiel$^1$, Alexandre Dauphin$^{2}$, Maciej Lewenstein$^{2,3}$, Jakub Zakrzewski$^{1,4}$, and Krzysztof Sacha$^{1,4}$} 

\address{$^1$Instytut Fizyki imienia Mariana Smoluchowskiego, 
Uniwersytet Jagiello\'nski, ulica Profesora Stanis\l{}awa \L{}ojasiewicza 11, PL-30-348 Krak\'ow, Poland}
\address{$^2$ICFO - Institut de Ciencies Fotoniques, The Barcelona Institute of Science and Technology, Av. Carl Friedrich Gauss 3, 08860 Castelldefels (Barcelona), Spain}
\address{$^3$ICREA, Pg. Llu\'is Companys 23, 08010 Barcelona, Spain}
\address{$^4$Mark Kac Complex Systems Research Center, Uniwersytet Jagiello\'nski, ulica Profesora Stanis\l{}awa \L{}ojasiewicza 11, PL-30-348 Krak\'ow, Poland
}

\begin{abstract}
By analogy with the  formation of space crystals, crystalline structures can also appear in the time domain. While in the case of space crystals we often ask about periodic arrangements of atoms in space at a moment of a detection, in time crystals the role of space and time is exchanged. That is, we fix a space point and ask if the probability density for detection of a system at this point behaves periodically in time. 
Here, we show that in periodically driven systems it is possible to realize topological insulators, which can be observed in time. The bulk-edge correspondence is related to the edge in time, where edge states localize. We focus on two examples: Su-Schrieffer-Heeger (SSH) model in time and Bose Haldane insulator which emerges in the dynamics of a periodically driven many-body system.    
\end{abstract}
\date{\today}

\maketitle
\ioptwocol

\section{Introduction} 
Time crystals are quantum systems where a crystalline structure emerges in time with no initial crystalline structure in space~\cite{Wilczek2012,Bruno2013b,Watanabe2015,Syrwid2017,Sacha2017rev} (for the classical version of time crystals, including topologically protected systems, see \cite{Shapere2012,Ghosh2014,Yao2018,Das2018,Alvarez2017,Aviles2017}). A crystalline structure in time is related to the periodic dynamics of a system. It has been shown that periodically driven quantum many-body systems can spontaneously self-reorganize their motion and start moving with a period different from the period of the driving \cite{Sacha2015,Khemani16,ElseFTC} (see also \cite{Yao2017,Lazarides2017,Russomanno2017,Zeng2017,Nakatsugawa2017,Ho2017,Huang2017,Gong2017,
Wang2017,Bomantara2018,Autti2018,Kosior2018,Mizuta2018,Giergiel2018a,Kosior2018a}). This kind of spontaneous formation of a new crystalline structure in time is dubbed discrete or Floquet time crystals and has been already realized experimentally~\cite{Zhang2017,Choi2017,Pal2018,Rovny2018,Rovny2018a}. However, periodically driven systems can also reveal a whole variety of condensed matter phases in the time domain even if no spontaneous process is involved in the emergence of such crystalline structures in time \cite{Sacha15a,sacha16,Giergiel2017,Mierzejewski2017,delande17,Giergiel2018,Giergiel2018a} (see \cite{Guo2013,Guo2016,Guo2016a,Liang2017} for phase space crystals). 
Indeed, if a single- or many-body system is driven resonantly, its resonant dynamics can be reduced to solid state-like behavior and importantly such condensed matter physics emerges not in the configuration space but in the time domain --- e.g. Anderson or many-body localization in time, superfluid-Mott insulator transition or quasi-crystals in the time domain have been demonstrated  \cite{Sacha15a,sacha16,Giergiel2017,Mierzejewski2017,delande17,Giergiel2018}.

In this paper we show that it is possible to drive resonantly a system so that the emerging crystalline structure in time is a symmetry protected topological (SPT) phase~\cite{Senthil_2015}. The topological time crystals we consider should not be confused with the so-called Floquet topological systems. In the latter, a crystalline structure (usually an optical lattice) is present in space and it is periodically driven so that its effective parameters can be changed and the system can reveal topological properties in space but no crystalline structure can be observed in time \cite{przysiezna15,Biedron16,Cardano2017,Maffei2018}. Our systems are also different from Floquet-Bloch systems where time periodicity is considered as an additional synthetic dimensional combined with a crystalline structure in space \cite{Kitagawa2010,Keyserlingk16a,Else16,Potter2016,Roy2016,Xu18}. We consider systems where external forces do not reveal any periodic behavior in space. They drive systems periodically in time and due to 
the resonant driving condensed matter phenomena emerge in dynamics of systems. 

In the models considered, resonant dynamics leads to an emergence of a time crystalline structure that may possess a SPT phase. SPT phases constitute a new paradigm: they are 
characterized by a global topological invariant and, therefore, can not be described by the Landau theory of phase transitions. Remarkable examples of these phases are, among others,  the Haldane phase~\cite{Haldane83,Haldane83a} and the topological insulators~\cite{Hasan2010}. The study of the latter with and without interactions has attracted much interest in condensed matter and in quantum simulators~\cite{Hasan2010,Goldman2014,Ozawa2018}. Quantum simulators constitute very versatile platforms with an unprecedented degree of control of the parameters of the system such as the hopping or the interactions~\cite{Lewenstein2007Book}. Quantum simulators have successfully simulated and detected topological insulators in 1D~\cite{Kitagawa2012,Cardano2017,Meier2016,Atala_2013,St_Jean17,
Meier2018,Leseleuc2018}, 2D~\cite{Aidelsburger2015,Stuhl2015,Tarnowski2017} and 4D~\cite{Lohse2018,Zilberberg2018}. All these realizations of the topological insulators rely on an initial underlying lattice. In 
the present paper we do not assume any underlining spatially periodic structure with a non-trivial topology~\cite{Bomantara2018}. We show that time crystals with topological properties can emerge due to an appropriate resonant driving and discuss possible implementation and detection schemes in quantum simulators.

\section{The model} 
\begin{figure}
\subfigure[]{\label{fspect1} 	            
\includegraphics[width=1.\columnwidth]{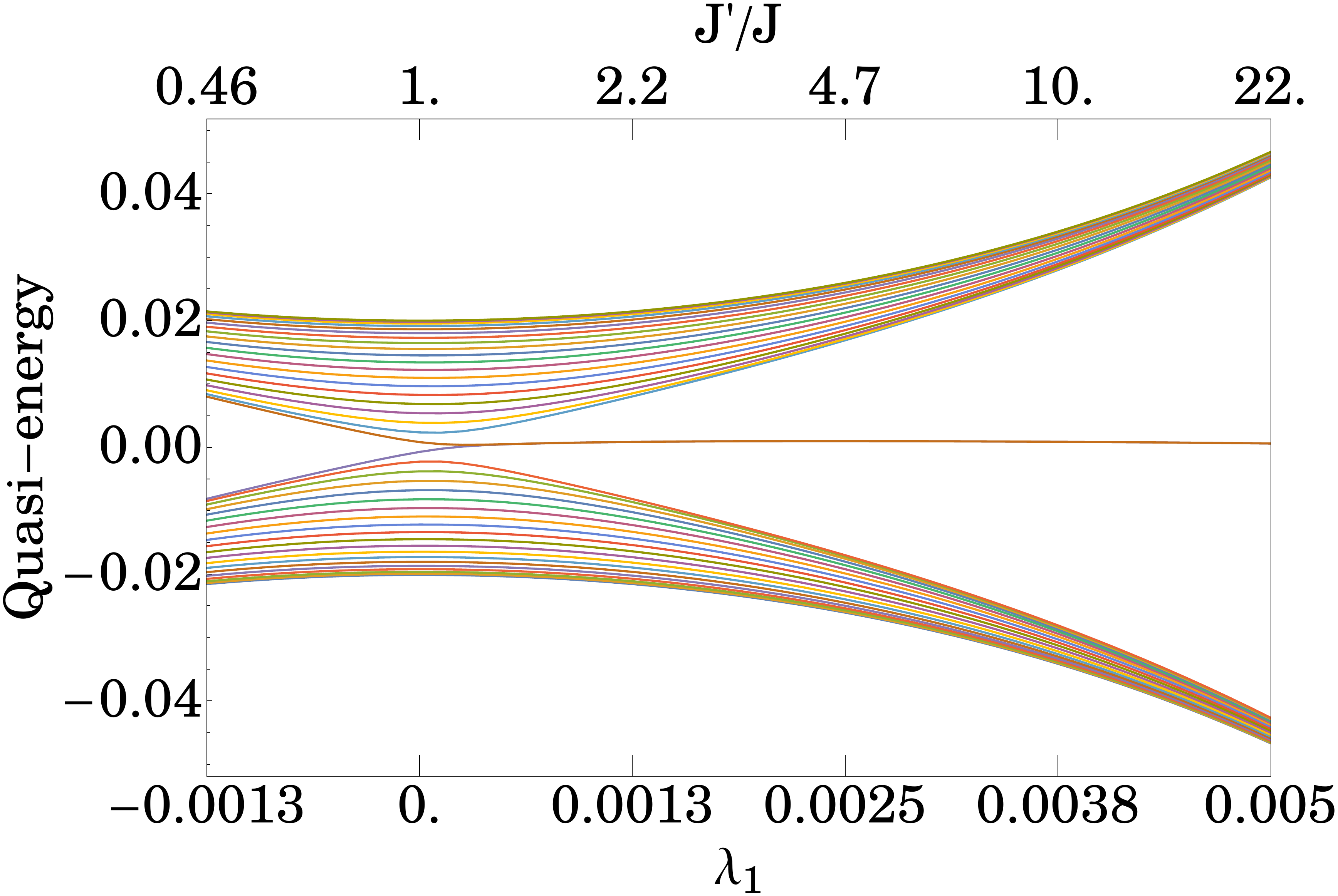}}
\subfigure[]{\label{fspect2} 
\includegraphics[width=1.\columnwidth]{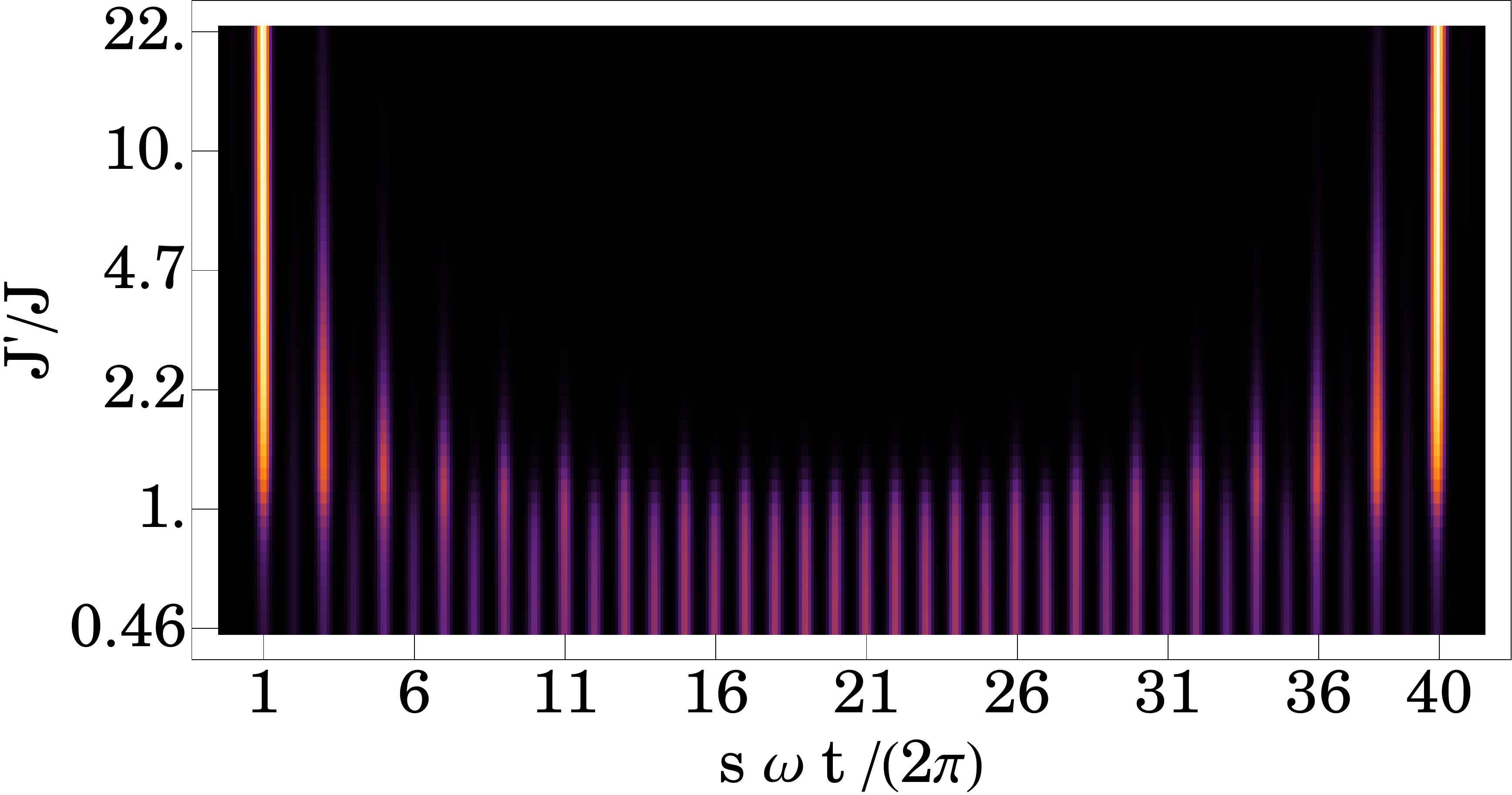}}       
\caption{SSH model in the time domain, i.e.  a particle bouncing on a periodically oscillating atom mirror, cf. (\ref{h}). Top panel shows the quasi-energy spectrum related to the $42:1$ resonant manifold ($s=42$) versus $\lambda_1$, or equivalently the ratio $J'/J$ of the tunneling amplitudes of the effective tight-binding Hamiltonian~(\ref{tbH}). At two adjacent sites of the tight-binding Hamiltonian~(\ref{tbH}) a barrier is created, i.e. $a_{s-1}=a_s=0$, which allows one to introduce the edge in the system. The barrier is realized by means of an additional modulation of the mirror motion $f(t)=\sum_{k}f_{k}e^{ik\omega t}$ where $f_k=\left(10^{-6}\right)k^2\omega^2\cos(k\pi/42){\rm sinc}^2(k\pi/121)$. {The zero energy value in the plot has been chosen so that it coincides with the average value of the spectrum presented. Note also that quasi-energy values are given in the gravitational units but they have been multiplied by $10^5$.}
Bottom panel presents one of the two numerical eigenstates corresponding to quasi-energies closest to zero, cf. top panel. In the topological phase, these eigenstates have zero quasi-energy and are linear combinations of the edge states localized at the edge of time. That is, in the laboratory frame, a 
detector is 
placed close to the oscillating mirror ($x\approx 0$) and the 
probability density of clicking of the detector is shown versus time for different values of the ratio $J'/J$ of the tunneling amplitudes in (\ref{tbH}). This behavior is repeated with the period $2\pi/\omega$. At $s\omega t/(2\pi)=1$ and $s\omega t/(2\pi)=40$ there are edges where the eigenstate localizes if $J'/J>1$.  The results correspond to $\omega=0.067$, $\lambda=0.06$ in (\ref{h}) and are obtained within the quantum secular approach \cite{Berman1977}, see \ref{appA}.
}
\label{fspect}   
\end{figure}

\begin{figure} 	            
\includegraphics[width=1.\columnwidth]{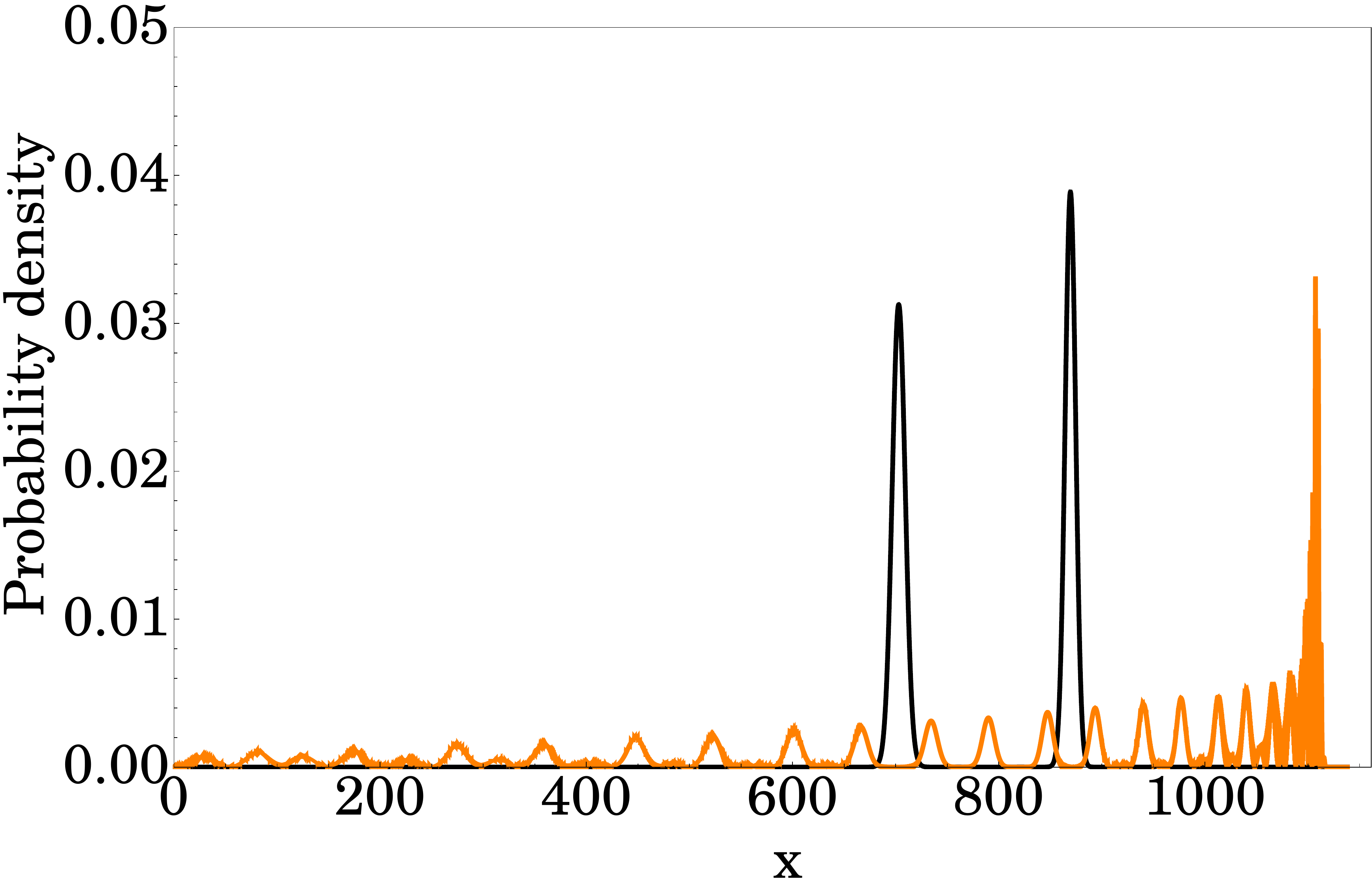}       
\caption{SSH model in the time domain. Black line: probability density of the eigenstate localized at the edge of time as shown in the bottom panel of Fig.~\ref{fspect} for $J'/J=22$ but plotted in the configuration space for $s\omega t=39\pi/2$. Orange line presents an example of a bulk eigenstate. The mirror is located at $x\approx 0$ while the classical turning point corresponds to $x=1110$. Note that the crystalline behavior described by Eq.~(\ref{tbH}) is not observed in the configuration space shown in this plot but in the time domain, see bottom panel of Fig.~\ref{fspect}.}
\label{edgespace}   
\end{figure}

We focus on ultra-cold atoms bouncing on an oscillating atom mirror in the presence of the gravitational field \cite{Steane95} (for the stationary mirror experiments see \cite{Roach1995,Sidorov1996,Westbrook1998,Lau1999,Bongs1999,Sidorov2002,Fiutowski2013,Kawalec2014}) but the phenomena we investigate can be realized in any periodically driven system which can reveal non-linear resonances in the classical description \cite{Lichtenberg1992}. The single-particle Hamiltonian, in the gravitational units and in the frame oscillating with the mirror \cite{Buchleitner2002}, reads $H=H_0+H_1$ where $H_0=p^2/2+x$ and 
\be
H_1=x\left[\lambda \cos(s\omega t)+\lambda_1\cos(s\omega t/2)+f(t)\right],
\label{h}
\ee 
with $f(t)=f(t+2\pi/\omega)=\sum_{k}f_{k}e^{ik\omega t}$ and integer-valued $s\gg 1$. Let us start with classical mech. description. In order to describe a resonant driving of the system it is convenient to perform canonical transformation to the so-called \emph{action angle variables}, where the unperturbed Hamiltonian depends only on the new momentum (action), $H_0(I)=(3\pi I)^{2/3}/2$ \cite{Lichtenberg1992,Buchleitner2002}. In the absence of the perturbation ($H_1=0$) the action $I$ is a constant of motion, and the conjugate position variable (angle) changes linearly in time $\theta(t)=\Omega t+\theta(0)$, where $\Omega(I)=dH_0(I)/dI$ is the frequency of periodic evolution of a particle. We assume the resonant driving $\Omega(I_0)=\omega$, where $I_0$ is the resonant value of the action. Then, by means of the secular approximation \cite{Lichtenberg1992,Buchleitner2002} (see \ref{appA})
 in the frame moving along the classical resonant orbit $\Theta=\theta-\omega t$, we obtain the effective Hamiltonian 
\bea
H_{\rm eff}=\frac{P^2}{2m_{\rm eff}}-V_0\cos(s\Theta)-V_1\cos(s\Theta/2)-V_2(\Theta), 
\label{sshH}
\eea
with $m_{\rm eff}=-\pi^2/\omega^4$, $V_0=\lambda(-1)^{s}/(s^2\omega^2)$, $V_1=4\lambda_1(-1)^{s/2}/(s^2\omega^2)$ and $V_2(\Theta)=\sum_ne^{in\Theta}(-1)^{n}f_{-n}/(n^2\omega^2)$. This result is valid in the regime close to resonance, i.e. when $P=I-I_0\approx 0$.  Equation~(\ref{sshH}) indicates that for $s\gg 1$ and $V_2(\Theta)=0$, a resonantly driven particle behaves like an electron in a one-dimensional (1D) space crystal. In the laboratory frame, the crystalline behavior described by Eq.~(\ref{sshH}) is reproduced in the time domain due to the linearity of the transformation $\Theta=\theta-\omega t$ \cite{Sacha15a,sacha16}. In other words, when we switch to the quantum description, the clicking probability of a detector located in the laboratory frame close to the classical resonant orbit, reflects periodic behavior of the Bloch waves $\psi(\Theta)$ of the Hamiltonian~(\ref{sshH}) in the moving frame. {In order to introduce an edge in the system and investigate the edge-bulk correspondence in a topological regime, we introduce a barrier $V_2(\Theta)$ localized on the sites $s-1$ and $s$ of the periodic potential in (\ref{sshH}). It is done by means of an additional modulation of the mirror motion $f(t)$ in (\ref{h}) whose Fourier components are $f_k=\left(10^{-6}\right)k^2\omega^2\cos(k\pi/42){\rm sinc}^2(k\pi/121)$ for the case of $s=42$ that is considered in Figs.\ref{fspect}-\ref{edgespace}.}

\section{A time crystal topological insulator} 
Let us focus on the first energy band of the quantum version of the effective Hamiltonian of Eq.~(\ref{sshH}) with $V_2(\Theta)=0$. The Wannier states $w_i(\Theta)$ of the first energy band 
\cite{Dutta2015}, which are localized in single sites of the periodic potential in Eq.~(\ref{sshH}), correspond to localized wavepackets $w_i(x,t)$ moving along the classical resonant orbit with the period $2\pi/\omega$.\footnote{Note that due to the negative effective mass $m_{\rm eff}$, the
  first energy band is the highest in energy.} For both non-vanishing $V_0$ and $V_1$, the effective periodic potential describes a Bravais lattice with a two-point basis. Restricting ourselves to the first energy band, i.e. the wavefunction of a particle is expanded as $\psi=\sum_{i=1}^sa_iw_i(\Theta)$, we obtain a tight-binding Hamiltonian 
\bea
H_{\rm eff}\approx -\frac12\sum_{i=1}^s J_i\left(a_{i+1}^*a_i+c.c.\right),
\label{tbH}
\eea
where $J_{2i}=J'$ and $J_{2i-1}=J$, which is identical to the SSH model~\cite{Su1979}. The latter describes spinless fermions hopping on a 1D-lattice with staggered hopping amplitudes. Changing the ratio $\lambda_1/\lambda$ in (\ref{h}), allows one to control the ratio $J'/J$. This effective Hamiltonian belongs to the BDI class of the periodic table of the topological insulators and superconductors~\cite{Chiu2016} and is characterized by a $\mathbb{Z}$ topological invariant, the winding number $\nu$. For an infinite system with $J'>J$ ($J'<J$), the system is in a topological (trivial) phase with winding number $\nu=1$ ($\nu=0$). For a finite system, the topological phase exhibits zero energy edge states protected by the topology of the bulk. The SSH model has been experimentally realized in quantum simulators and both the presence of edge states and the winding number have been measured \cite{Atala_2013,St_Jean17,Meier2016}. Let us emphasize that the Wannier states $w_i(x,t)$ of Eq.~(\ref{sshH}) are 
localized wavepackets of the effective SSH Hamiltonian of Eq.~(\ref{tbH}). We then discuss how such states allow one to detect the topology of the SSH Hamiltonian.

\section{Detection of the topology} 
The experimental detection of the edge states and the corresponding bulk winding number $\nu$ can be realized in ultra-cold atoms which form a Bose-Einstein condensate (BEC) bouncing on an oscillating atom mirror. The Hamiltonian of Eq.~(\ref{tbH}) fulfills periodic boundary conditions, i.e. $a_{s+1}=a_1$. However, it is possible to introduce an edge in our system by means of a proper modulation of the mirror motion. Indeed, if $f_k$'s in the definition of $f(t)$ are suitably chosen, the resulting effective potential $V_2(\theta)$ can have a 
shape of a barrier localized on two adjacent sites of the periodic potential in Eq.~(\ref{sshH}). Then, the edge is created and two eigenstates can localize  exponentially close to it in the topological phase. The corresponding quasi-energy spectrum of the full Floquet Hamiltonian ${\cal H}(t)=H-i\partial_t$ is shown in Fig.~\ref{fspect1}. For $J'/J>1$ two degenerate zero-energy levels form which correspond to two eigenstates localized close to the barrier created by the potential $V_2(\Theta)$. In the laboratory frame these two eigenstates are related to Floquet states that evolve periodically along the classical resonant orbit. For a detector located close to the orbit, the probability of a detector clicking reveals an edge in time and these two Floquet states localize close to it, as shown in Fig.~\ref{fspect2} for increasing values of $J'/J$. In Ref.~\cite{Giergiel2018a} it {was proposed} how a BEC of non-interacting atoms can be loaded from a trap on a classical resonant orbit: a localized atomic cloud has to 
be released from 
a trap at the position of the turning point of the classical resonant orbit above the oscillating  mirror. Then, the initial state of the system corresponds to all atoms occupying a single localized state $w_i(x,t)$ which evolves along the orbit. An atomic cloud loaded when the edge state is passing close to the turning point remains close to the edge as the edge states do not penetrate much the bulk~\cite{Meier2016}. How one of the zero-energy eigenstates with the localization close to the edge in time looks like  in the configuration space is depicted in Fig.~\ref{edgespace} for $s\omega t=39\pi/2$. Measurements of atomic density at different times therefore allow one to confirm the localization properties of the state loaded on the edge. 

On the other hand, if an atomic cloud is released in the bulk, then the subsequent time evolution leads to non-zero populations $|a_i(t)|^2$ of many Wannier wavepackets. Measurement of the atomic density allows one to obtain $|a_i(t)|^2$ and consequently the winding number which is determined by the mean chiral displacement, i.e. $\nu\approx2\sum_i(i-i_0)(|a_{2i+1}(t)|^2-|a_{2i}(t)|^2)$ where $i_0$ is a number of a cell of the Bravais lattice where the atomic cloud is initially loaded \cite{Cardano2017,Maffei2018,Meier2018}. The relation between $\nu$ and the mean chiral displacement is valid after a long-time evolution when time averaged~\cite{Meier2018}.

\section{A time crystal Haldane phase} 
We now switch from single-particle to many-body systems which are resonantly driven and which can be characterized by non-trivial topology. It is known that a Bose gas  in a time-independent spatially periodic lattice with repulsive on-site and nearest-neighbor interactions, described by the following Bose-Hubbard Hamiltonian
\bea
\hat H&=&-\frac{J}{2}\sum_{i}\left(\hat a_{i+1}^\dagger\hat a_{i}+h.c.\right)+\frac{U}{2}\sum_{i}\hat n_{i}(\hat n_{i}-1) \cr && +V\sum_i \hat n_{i+1}\hat n_i,
\label{bhh}
\eea
can reveal a topological behavior. For large $J$ the Hamiltonian (\ref{bhh}) describes superfluid phase of bosons, for large $U$ the Mott insulator (MI) emerges and for large $V$ the density wave (DW) phase is present where translation symmetry of the Hamiltonian (\ref{bhh}) is spontaneously broken. However, between the MI phase and the DW phase there is the topological Haldane  insulator (HI) phase \cite{Torre2006,Rossini2012,Ejima2015,Biedron2018} -- or more precisely  a bosonic analog of HI in spin-1 chain~\cite{Haldane83,Haldane83a}. As discussed in Ref.~\cite{Torre2006} this phase breaks a hidden $Z_2$ symmetry related to a highly nonlocal string order parameter \cite{Kennedy92}.

\begin{figure} 	            
\includegraphics[width=1.\columnwidth]{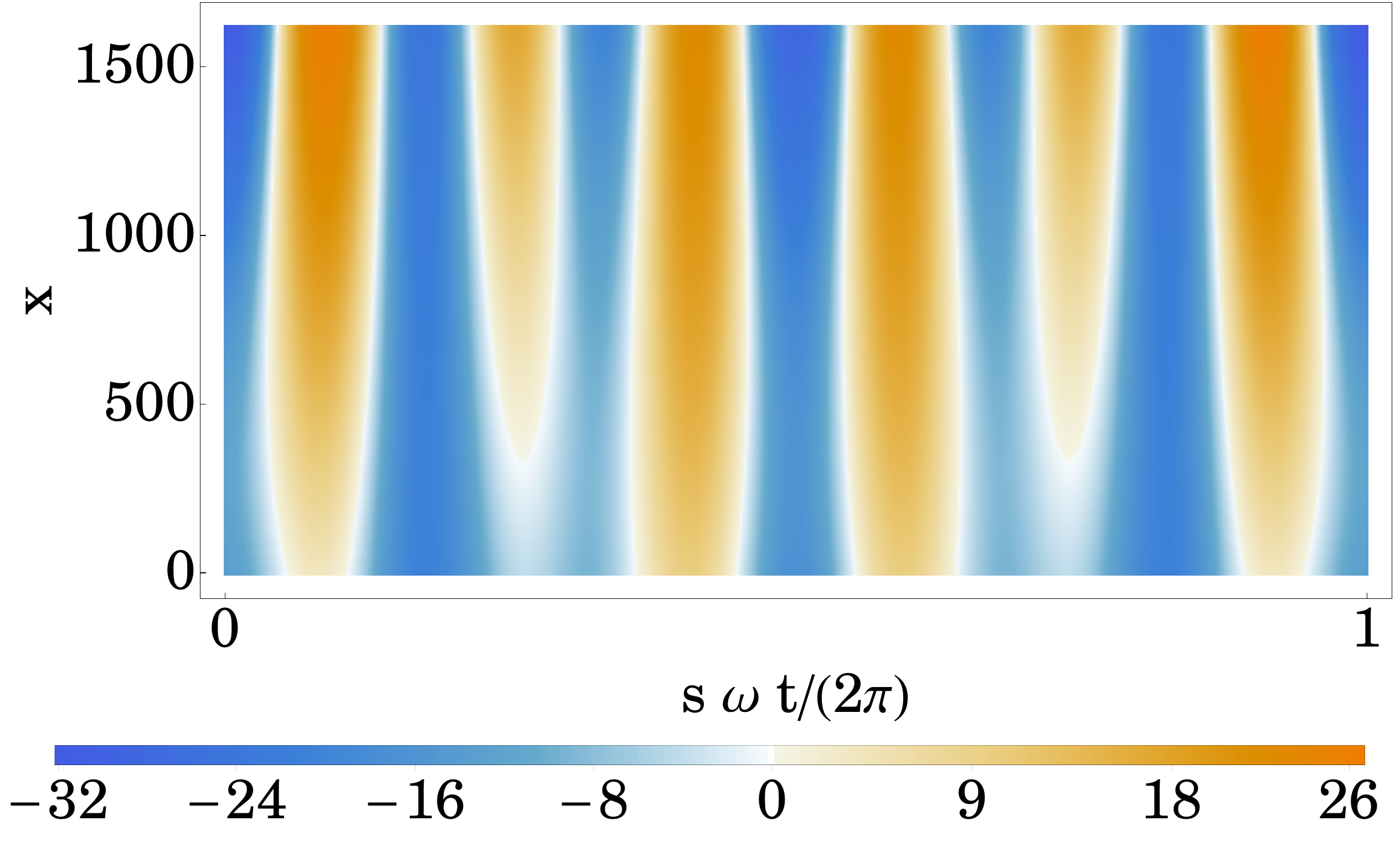}       
\caption{Ultra-cold atoms bouncing on a harmonically oscillating mirror, i.e. $\lambda_1=0$ and $f(t)=0$ in (\ref{h}). Figure shows an example of the dependence of the atomic s-wave scattering length $g_0(x,t)\times 10^{-4}$ on a position in space and on time which leads to the effective interactions between bosons described by the on-site and nearest neighbor interaction coefficients $U/J=3$ and $V/J=2.5$. If unit boson filling factor in the effective Bose-Hubbard model is assumed, then the results are related to the HI phase \cite{Ejima2015}. To realize MI or DW phases, the appropriate $g_0(x,t)$ is similar with the amplitude of its oscillations around zero being twice smaller or greater by at least factor 1.5, respectively. The other parameters of the system are the following: $\omega=0.056$, $s=64$ (i.e. $64:1$ resonance), $\lambda=0.14$ and $M=2$ in the expansion $g_0=\sum_{m=0}^M\alpha(t,m)x^m$. The mirror is located at $x=0$ and the turning point of the classical resonant orbit at $x\approx 1600$. The 
maximal temporal interaction energy per particle is about $1.6\times10^4J$ and thus much smaller than the gap between the first and second quasi-energy bands of the corresponding single-particle problem which is $1.1\times 10^5J$ where $J=5.5\times 10^{-8}$. The results are obtained within the quantum secular approach \cite{Berman1977}, see \ref{appA}.}
\label{g0tm}   
\end{figure} 

To realize a resonantly driven system that, in the time-periodic Wannier-like modes $w_i(x,t)$ of Eq.~(\ref{sshH}), is effectively described by the Hamiltonian of~(\ref{bhh}), we again consider ultra-cold bosons bouncing on an oscillating mirror. We assume that the single-particle Hamiltonian corresponds to Eq.~(\ref{h}) with $\lambda_1=0$ and $f(t)=0$. For the resonant driving described above, we may restrict to the first energy band of the effective Hamiltonian (\ref{sshH}) that leads to the tight-binding model~(\ref{tbH}). In the many-body case when we restrict to the Hilbert subspace spanned by Fock states $|n_1,\dots,n_s\ra$, where $n_i$'s denote numbers of bosons occupying Wannier wavepackets $w_i(x,t)$, we obtain the effective many-body Hamiltonian which resembles (\ref{bhh}) but with the interaction terms ${1/2}\sum_{i,j=1}^sU_{ij}\hat a_i^\dagger\hat a_j^\dagger\hat a_j\hat a_i$ where $U_{ij}=2\int_0^{2\pi/\omega}dt\int_0^\infty dx g_0|w_i|^2|w_j|^2$ for $i\ne j$ 
and similar $U_{ii}$ but twice smaller \cite{Sacha15a} (see \ref{appA}). The 
effective interaction coefficients $U_{ij}$ depend on the atomic s-wave scattering length $g_0$, shapes of  $|w_i(x,t)|^2$ and how densities of different Wannier wavepackets overlap in the course of time evolution on the classical resonant orbit. Despite the fact that the original interactions between ultra-cold atoms are contact, the effective interactions can be long-range \cite{Anisimovas2015,Sacha15a,Guo2016}. Moreover, they can be controlled by changing the s-wave scattering length in space and periodically in time by means of a Feshbach resonance, i.e. $g_0=g_0(x,t)$. Indeed, if the applied magnetic field results in appropriate oscillations of the scattering length around zero, nearly arbitrary effective long-range interactions can be created. In order to perform a systematic analysis of the control of $U_{ij}$, we assume that $g_0(x,t)=\sum_{m=0}^{M}\alpha(t,m)x^m$ and write the interaction coefficients in the form $U_{ij}=\sum_{m=0}^{M}\int_0^{2\pi/\omega}dt\alpha(t,m)u_{ij}(t,m)$. To find a suitable 
$\alpha(t,m)$ one can apply the 
singular 
value decomposition of the matrix $u_{ij}(t,m)$ where $(i,j)$ and $(t,m)$ are treated as indices of rows and columns, respectively \cite{Giergiel2018}. Left singular vectors tell us which sets of interaction coefficients $U_{ij}$ can be realized, while the corresponding right singular vectors give the recipes for $\alpha(t,m)$ and consequently for $g_0(x,t)$. In Fig.~\ref{g0tm} we show an example of $g_0(x,t)$ corresponding to $U_{ii}=U$, $U_{ij}=V$ for $|i-j|=1$ and $U_{ij}=0$ when $|i-j|>1$ where $U/J=3$ and $V/J=2.5$. These values are related to the Haldane insulator phase if the unit mean boson filling factor is assumed \cite{Ejima2015}. In order to realize topologically trivial MI or DW phases, the corresponding $g_0(x,t)/J$  may take a similar form with the amplitude of its oscillations around zero being about twice smaller for MI or at least a factor 1.5 greater for the DW phase. 
For example, for the parameters of Ref.~\cite{Giergiel2018a}, if one wants to modulate the scattering length around zero value with the amplitude of two Bohr radius, the magnetic field has to change with the speed 0.5G/ms --- well within the experimental reach \cite{Mark2018}.

It is quite straightforward to prepare initially all bosons in a single Wannier state $w_i(x,t)$ by releasing atomic cloud from a trap at the classical turning point above the mirror \cite{Giergiel2018a}. However, the preparation of the system in lowest quasi-energy state within the resonant subspace, i.e. in the ground state of (\ref{bhh}) is not easy. An open question remains whether the time-evolution after a quench of a localized initial state can reveal the topology of the Haldane Insulating phase. This problem is beyond the scope of the present paper.

{Finally let us analyse whether the resonant driving does not heat the system considerably. Consider first a single particle problem. While a harmonic oscillator can absorb continuously energy when resonantly driven, we consider a model with nonlinear resonances ---  a particle bouncing resonantly on an oscillating mirror  \cite{Buchleitner2002}.
For such a system its period of motion depends on energy, so  when a particle absorbs energy, its period changes and a particle gets out of the resonance. The resonant transfer of energy stops \cite{Lichtenberg1992}. The non-linear resonances are related to the presence of elliptical islands in the classical phase space \cite{Lichtenberg1992}, see Fig.~\ref{comps} in \ref{appA}. Floquet states located inside the islands  are strongly localized \cite{Buchleitner2002}. Their coupling to other states localized outside the islands, for any type of weak additional coupling is exponentially small precisely due to the semiclassical localization. Such a coupling may be provided by e.g. weak interactions among the particles that require a consideration of necessarily enlarged $2N$-dimensional phase space. Such an additional coupling of a subspace of  Floquet states localized in the islands to the complementary Hilbert subspace can be tested within the mean field approach as an exact many-body analysis is obviously beyond the current possibilities. A careful mean-field analysis performed in~\cite{Giergiel2018a} indicates that, on the relevant experimentally time scale, such a coupling (and the resulting heating) is negligible also if typical experimental imperfections are taken into account. Thus the relevant physics takes place in the resonant
subspace of Floquet states localized in the phase space and are immune to heating, at least for a time scale much longer than the experimentally relevant one \cite{Giergiel2018a}, without any disorder present.
If interactions are sufficiently weak then the physics may be described within the lowest band of the Bose-Hubbard model constructed in the paper.}

\section{Discussion}
We have shown how one can construct topological insulators in the time domain. To this end we considered bosons bouncing on the periodically oscillating atom mirror, topological phases are obtained by an appropriate tuning the shape of the oscillations. We presented explicit schemes for the realization of the effective SSH and the extended Bose-Hubbard Hamiltonians which possess topological behavior in the time domain. In the case of the topological SSH, we proposed a detection method of the topological properties in time with state-of-the-art  techniques~\cite{Giergiel2018a}. In the case of the Haldane insulator, the experimental detection of the topological phase remains a challenge.

{\it Note added:} After the submission of the present manuscript we learned about the work on a similar topic but in photonic materials \cite{Lustig2018}.

\section*{Acknowledgments} AD and ML acknowledge fundings from the Spanish Ministry MINECO (National Plan
15 Grant: FISICATEAMO No. FIS2016-79508-P, SEVERO OCHOA No. SEV-2015-0522, FPI), European Social Fund, Fundaci\'o Cellex, Generalitat de Catalunya (AGAUR Grant No. 2017 SGR 1341 and CERCA/Program), ERC AdG OSYRIS and EU FETPRO QUIC.
AD is financed by a Cellex-ICFO-MPQ fellowship. Support of the National Science Centre, Poland via Projects No. 2016/20/W/ST4/00314 (K.G., M.L.), 2016/21/B/ST2/01086 (J.Z.), and No. 2016/21/B/ST2/01095 (K.S.) is acknowledged.

\appendix


\section{}
\label{appA}

In this Appendix we present the derivation of the effective Hamiltonians used in the paper. We also provide a detailed explanation of the emergence of crystalline structures in the time domain in the course of dynamical evolution of resonantly driven systems.

\subsection{ Classical effective Hamiltonian}

We consider a single atom bouncing on an oscillating atom mirror in the presence of the gravitational force. The Hamiltonian of the system, in the frame oscillating with the mirror \cite{Buchleitner2002} (then the mirror is fixed) and in the gravitational units, reads
\bea
H&=&H_0+H_1, \label{hfulls}\\
H_0&=&\frac{p^2}{2}+x, \label{h0fulls}\\
H_1&=&x\left[\lambda \cos(s\omega t)+\lambda_1\cos(s\omega t/2)+f(t)\right], \label{h1fulls}
\eea
where $H_1$ describes the oscillation of the mirror  composed of two harmonics. $s$ is an integer number and
\be
f(t)=f(t+2\pi/\omega)=\sum_{k}f_{k}e^{ik\omega t}.
\ee
 We assume $x\ge 0$ with the mirror located at $x=0$. At first glance the system does not have anything in common with a solid state-like behavior. In order to demonstrate that for a resonant driving of the particle motion, the system behaves like a condensed matter system (and in particular may possess topological properties) let us begin with the classical description and apply the secular approximation \cite{Lichtenberg1992,Buchleitner2002}. 

First we apply the canonical transformation from the Cartesian position $x$ and momentum $p$ to the so-called action-angle variables. Then, the unperturbed Hamiltonian depends on the new momentum (action) only \cite{Buchleitner2002},
\be
H_0=\frac12(3\pi I)^{2/3},
\ee
and the solution of the unperturbed particle motion is straightforward. That is, the action is a constant of motion, $I=$const, and the conjugate position variable (the angle) evolves linearly in time, $\theta(t)=\Omega(I)\; t+\theta(0)$, where $\Omega(I)=dH_0(I)/dI$ is the frequency of periodic evolution of a particle. The perturbation part of the Hamiltonian, in the new canonically conjugate variables, reads
\bea
H_1&=&\left(\sum_n h_n(I)e^{in\theta}\right)\left(\sum_mF_me^{im\omega t}\right),
\eea
where $h_0=\left(\frac{\pi I}{\sqrt{3}}\right)^{2/3}$ and $h_n=\frac{(-1)^{n+1}}{n^2}\left(\frac{3I}{\pi^2}\right)^{2/3}$ for $n\ne 0$, and $F_m$ are Fourier components of the time-periodic function in (\ref{h1fulls}).

\begin{figure} 	            
\includegraphics[width=1.\columnwidth]{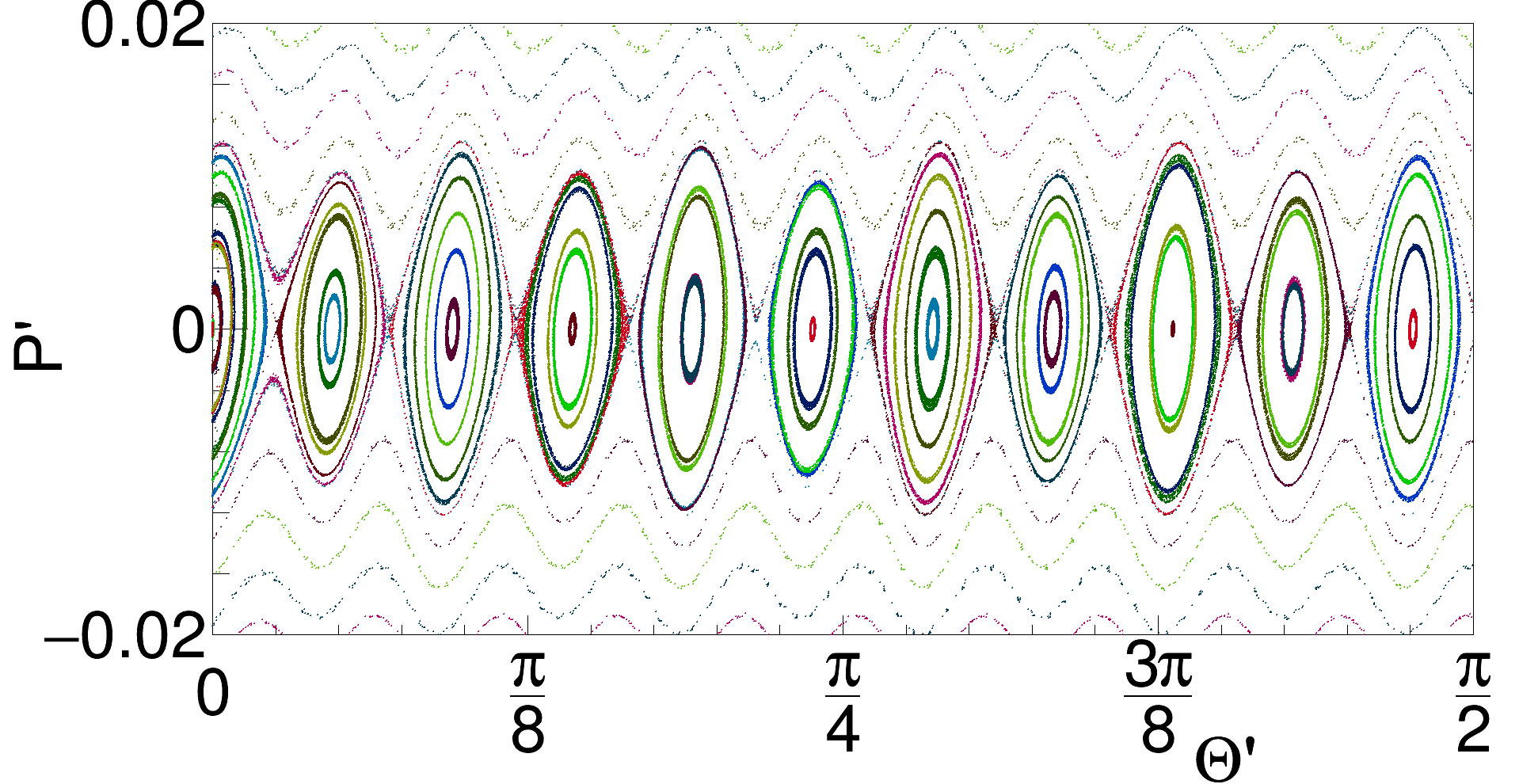}       
\includegraphics[width=1.\columnwidth]{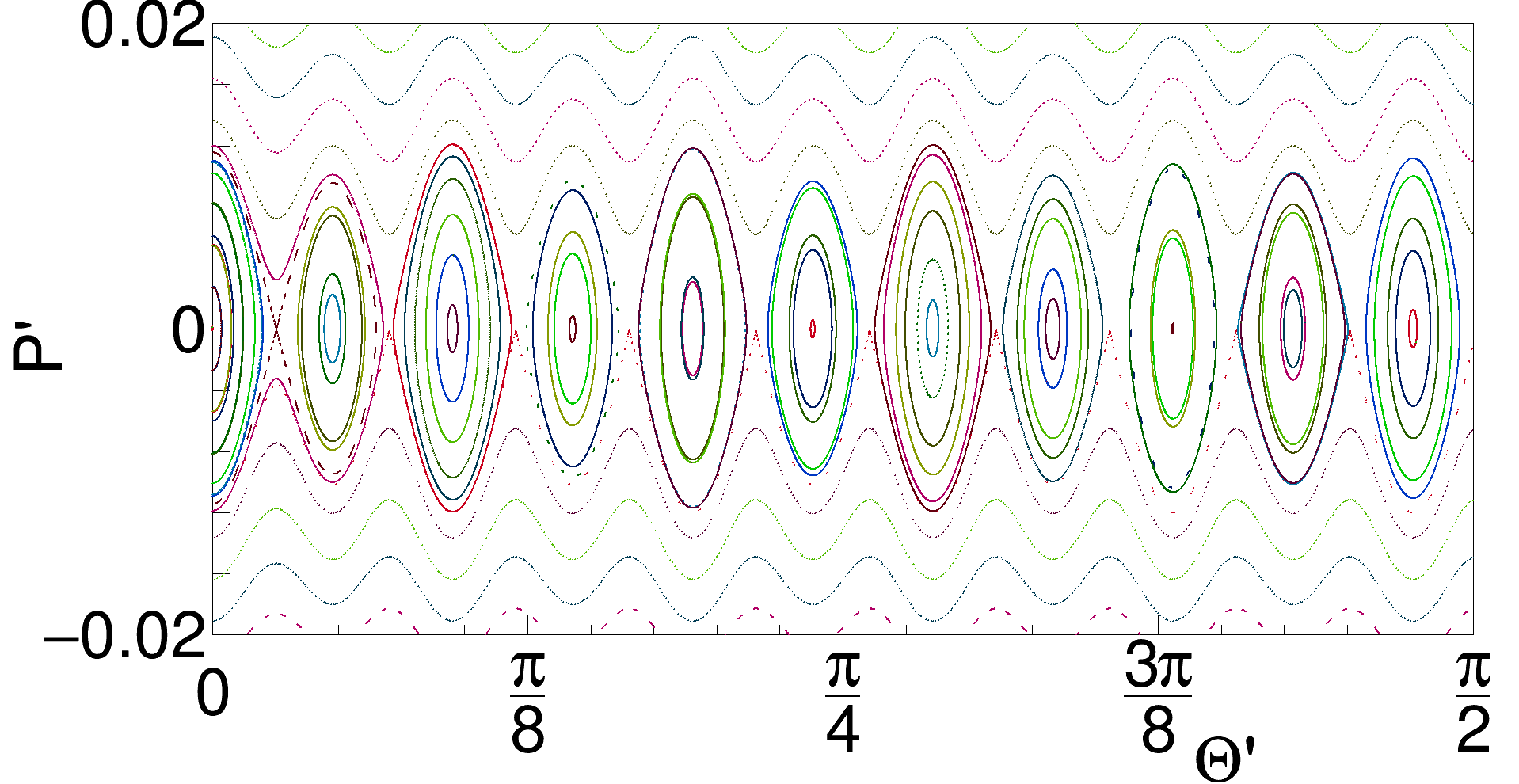}       
\caption{
Comparison of the exact stroboscopic picture (top panel) of the system phase space with the phase space portrait (bottom panel) obtained with the help of the effective Hamiltonian (\ref{sshHs}). The parameters of the system are the same as those used in the paper in Figs.~1-2, i.e. $s=42$, $\omega=0.067$, $\lambda=0.06$, $\lambda_1=0.005$ and $f_k=\left(10^{-6}\right)k^2\omega^2\cos(k\pi/42){\rm sinc}^2(k\pi/121)$. Scaled variables have been used in the plot, i.e. $P'=P/I_0$ and $\Theta'=\Theta$. Note that for the
sake of clarity we show only a quarter of the full $2\pi$ range of $\Theta'$.
}
\label{comps}   
\end{figure} 

Next, we assume that a particle is resonantly driven, that is the frequency $\omega$ is equal to the frequency of the unperturbed motion, i.e. $\omega=\Omega(I_0)$ where $I_0$ is the resonant value of the action. We are interested in initial conditions of a particle where the action $I\approx I_0$. Then, it is convenient to perform another canonical transformation to the moving frame, associated with the  motion along the resonant orbit \cite{Lichtenberg1992},
\bea
\Theta&=&\theta-\omega t, \label{thlins} \\
{\cal I}&=&I.
\eea
In that frame, the Hamiltonian of the system takes the form
\bea
H&=&\frac12\left(3\pi {\cal I}\right)^{2/3}-\omega {\cal I} 
+\sum_{n,m}h_n({\cal I})F_me^{in\Theta}e^{i(n+m)\omega t}. \cr &&
\label{fullmovHs}
\eea
Now, if we are interested in the motion close to the resonant orbit, i.e. when ${\cal I}\approx I_0$, both $\cal I$ and $\Theta$ are changing slowly if the perturbation is weak. The only fast variable is time and we can significantly simplify the description by averaging the Hamiltonian (\ref{fullmovHs}) over time \cite{Lichtenberg1992} that results in 
\bea
H_{\rm eff}&=&\frac{P^2}{2m_{\rm eff}}-V_0\cos(s\Theta)-V_1\cos(s\Theta/2)-V_2(\Theta), \cr &&
\label{sshHs}
\eea
which is the effective Hamiltonian (2) in the paper. In~(\ref{sshHs}) 
\bea
m_{\rm eff}&=&-\frac{\pi^2}{\omega^4}, \\
V_0&=&\lambda\frac{(-1)^{s}}{s^2\omega^2}, \\
V_1&=&\lambda_1\frac{4(-1)^{s/2}}{s^2\omega^2}, \\
V_2(\Theta)&=&\sum_n\frac{(-1)^{n}}{n^2\omega^2}f_{-n}e^{in\Theta},
\eea
and we have performed Taylor expansion around $P={\cal I}-I_0\approx 0$.
The validity of the effective Hamiltonian (\ref{sshHs}) can be easily tested by comparing the exact stroboscopic picture of the system phase space with the phase space portrait generated by the Hamiltonian (\ref{sshHs}). It is done in Fig.~\ref{comps} which demonstrates that the secular approach captures quantitatively the dynamics of the system.

\subsection{Quantum effective Hamiltonian}

So far we have performed the classical analysis. In order to switch to the effective quantum description we can either perform a quantization of the classical effective Hamiltonian (\ref{sshHs}) or apply a quantum version of the secular approximation for the Hamiltonian (\ref{hfulls}) \cite{Berman1977}. Let us first discuss the former approach. Classical equations of motion possess the scaling symmetry which implies that by a proper rescaling of the parameters and the dynamical variables of the system we obtain the same behavior as presented in Fig.~\ref{comps} but around arbitrary value of $I_0$ \cite{Buchleitner2002}. That is, when we redefine $\omega'=I_0^{-1/3}\omega$ and $\lambda'=\lambda$ we can use the results presented in Fig.~\ref{comps} if we rescale $p'=I_0^{1/3}p$, $x'=I_0^{2/3}x$ and $t'=I_0^{1/3}t$. In the quantum description the scaling symmetry is broken because the Planck constant sets a scale in the phase space,
\be
[x,p]=i \quad \Rightarrow \quad [x',p']=\frac{i}{I_0}.
\label{scalings}
\ee
From (\ref{scalings}) we see that $1/I_0$ plays a role of the effective Planck constant. Thus, for $I_0\gg 1$, quantization of the effective Hamiltonian (\ref{sshHs}), obtained in the action-angle variables, i.e. when $P\rightarrow -i\frac{\partial}{\partial \Theta}$, provides perfect quantum description of the resonant behavior of the system. The same quantum results can be obtained by applying the quantum secular approach \cite{Berman1977} which yields
\be
\la n'|H_{\rm eff}|n \ra=\left(E_n-n\omega\right)\delta_{nn'}+\la n'|x|n\ra F_{n-n'},
\label{sqeffh}
\ee
where $|n\ra$'s are eigenstates of the unperturbed system and $2^{1/3}E_n$ are zeros of the Airy function \cite{Buchleitner2002} and $F_n$ are Fourier components of the time periodic function in (\ref{h1fulls}). Equation~(\ref{sqeffh}) has been obtained by switching to the moving frame, with the help of the unitary transformation $\hat U=e^{i\hat n\omega t}$, and by averaging the resulting Hamiltonian over time \cite{Berman1977}. 

\subsection{Crystalline structure in time}

For $f(t)=0$, and consequently for $V_2(\Theta)=0$ in (\ref{sshHs}), and for $s\gg 1$, corresponding classically to the $s:1$ resonance between the motion of the particle in the gravitational field and mirror oscillations, the effective Hamiltonian becomes identical to a solid state system with an electron moving in a periodic potential in space with periodic boundary conditions. In the quantum description, eigenstates of (\ref{sshHs}) are Bloch waves $e^{ik\Theta}u_n(\Theta)$ where $k$ is a quasi-momentum, $n$ denotes a number of an energy band and $u_n(\Theta+2\pi/s)=u_n(\Theta)$. Such a crystalline structure that we observe in the frame moving along the resonant orbit is reproduced in the time domain when we return to the laboratory frame \cite{Sacha15a,sacha16}. Indeed, when we locate a detector in the laboratory frame close to the resonant orbit (i.e. at a space point where $I\approx I_0$ and $\theta=$const), probability for clicking of the detector as a function of time reproduces periodic behavior 
versus $\Theta$ that we obtain in the moving frame with the help of the effective Hamiltonian (\ref{sshHs}). This is the crucial property of the system we consider here and it is related to the fact that the transformation between the laboratory frame and the moving frame is linear in time, see (\ref{thlins}). Thus, for $\theta=$const we obtain the same behavior in time as versus $\Theta$ in the moving frame, $e^{ik(\theta-\omega t)}u_n(\theta-\omega t)=e^{ik\Theta}u_n(\Theta)$. We would like to stress that such a crystalline structure in time is not a result of the presence of any potential periodic in space. It emerges in the dynamics of the system due to the resonant driving and a high resonance order $s$.

\subsection{Tight-binding approximation}

Assume that $V_2(\Theta)=0$ in (\ref{sshHs}).
If we are interested in the first energy band of the effective Hamiltonian (\ref{sshHs}), the description of the system can still be simplified. Suppose that $w_i(\Theta)$ are Wannier functions corresponding to the first energy band of (\ref{sshHs}). Then, expanding the wavefunction of a particle in the Wannier function basis, $\psi(\Theta)=\sum_{i=1}^sa_iw_i(\Theta)$, we obtain the tight-binding Hamiltonian, cf. (3) in the paper,
\be
H_{\rm eff}\approx -\frac12\sum_{i=1}^s J_i\left(a_{i+1}^*a_i+c.c.\right),
\label{tbHs}
\ee
where tunneling amplitudes $J_i=-2\int_0^{2\pi}d\Theta w_{i+1}^*H_{\rm eff}w_i$ describe tunneling of a particle between neighbouring sites of the potential in (\ref{sshHs}). There are also longer range tunnelings but they are orders of magnitude weaker for the parameters we consider in the paper. For $\lambda_1\ne 0$ when $J=J_{2i-1}$, $J'=J_{2i}$ and $J\ne J'$, the Hamiltonian (\ref{tbHs}) describes a particle in a Bravais lattice with a two-point basis. Actually it is the SSH model \cite{Su1979} which can possess topological properties.

The tight-binding Hamiltonian (\ref{tbHs}) describes a particle in the Wannier function basis. In the laboratory frame this is the basis of localized wavepackets which evolve periodically along the resonant orbit, i.e. in the Cartesian coordinate frame $w_i(x,t+2\pi/\omega)=w_i(x,t)$. It shows that condensed matter physics in the time domain we consider here can be described by a solid state model but in the time-periodic basis.

\subsection{Many-body problem}

Having introduced a tight-binding model for a single particle problem, we can easily switch to the many-body case. Indeed, assume that we have $N$ bosons which interact via Dirac-delta potential $g_0\delta(x)$ and we restrict to the resonant Hilbert subspace of the system. That is, we restrict to the Hilbert subspace spanned by Fock states $|n_1,\dots,n_s\ra$ where $n_i$ denotes a number of particles that occupy a  Wannier state $w_i(x,t)$. Then, the effective many-body Hamiltonian takes the form of the Bose-Hubbard model \cite{Sacha15a},
\bea
\hat H&=&-\frac{1}{2}\sum_{i}\left(J_i\hat a_{i+1}^\dagger\hat a_{i}+h.c.\right)+\frac12\sum_{i,j=1}^sU_{ij}\hat a_i^\dagger\hat a_j^\dagger\hat a_j\hat a_i, \cr &&
\label{bhhs}
\eea
where $\hat a_i$ are standard bosonic annihilation operators and   
\bea
U_{ij}&=&2\int\limits_0^{2\pi/\omega}dt\int\limits_0^\infty dx \;g_0\;|w_i(x,t)|^2|w_j(x,t)|^2, \quad {\rm for} \; i\ne j, \cr 
U_{ii}&=&\int\limits_0^{2\pi/\omega}dt\int\limits_0^\infty dx \;g_0\;|w_i(x,t)|^4,
\eea
are coefficients which describe interactions between particles. If $n_i$ bosons occupy a given localized Wannier-like wavepacket $w_i(x,t)$ they interact that is related to the on-site interactions in (\ref{bhhs}). However, bosons occupying different localized wavepackets $w_i(x,t)$ and $w_j(x,t)$ also interact because in the course of time evolution along the resonant orbit the wavepackets pass each other. That leads to effective long-range interactions in the Bose-Hubbard Hamiltonian (\ref{bhhs}). Moreover, the long-range interaction can be engineered if the s-wave scattering length of atoms is periodically modulated in time and depends on a position in space, i.e. $g_0=g_0(x,t)$.

The effective Bose-Hubbard Hamiltonian is valid if the maximal interaction energy per particle is much smaller than the energy gap {$\Delta E$} between the first and second energy bands of the single-particle Hamiltonian (\ref{sshHs}). {In the paper we assume that the interaction energy is larger than the width of the first energy band (which is of the order of $J$) but much smaller than $\Delta E$. There is no problem to fulfill this condition because $\Delta E\gg J$.}



\section*{References}
\providecommand{\newblock}{}


\end{document}